\documentclass[journal]{IEEEtran}

\usepackage{pifont,calc}
\usepackage{cite}

\usepackage{graphicx}
\ifCLASSINFOpdf
   \usepackage[pdftex]{graphicx}
\else
 \fi
\usepackage[cmex10]{amsmath}
\usepackage{array}

\usepackage{cases} % to combine equations together
\usepackage{bm} % to make xila zimu hei ti

\makeatletter

\newcommand{\Rmnum}[1]{\expandafter\@slowromancap\romannumeral #1@}
\makeatother
\usepackage{booktabs}

\usepackage[Symbol]{upgreek}

\begin{document}

\title{Comments on ``A New ML Based Interference Cancellation Technique for Layered
Space-Time Codes"}

\author{Hufei Zhu,~\IEEEmembership{Member,~IEEE,}
        Wen Chen,~\IEEEmembership{Member,~IEEE}
       %Bin Li,~\IEEEmembership{Member,~IEEE}
        % <-this % stops a space
\thanks{Hufei Zhu is with Department of Electronic Engineering, Shanghai Jiao Tong University,
Shanghai 200240, P. R. China and Huawei Technologies Co., Ltd.,
Shenzhen 518129, P. R. China, e-mail: zhuhufei@yahoo.com.cn;}
\thanks{Wen Chen is with Department of Electronic Engineering,
Shanghai Jiao Tong University, Shanghai 200240, P.R. China and SEU
SKL for Mobile Communications, e-mail: wenchen@sjtu.edu.cn.}
\thanks{This work is supported by NSF China \#60972031, by SEU SKL project
\#W200907, by Huawei Funding \#YJCB2009024WL and \#YJCB2008048WL, by
New Century Excellent Talent in Universities \#NCET-06-0386, and by
National 973 project \#2009CB824900.
%the Department of Wireless Research, Huawei Technologies Co., Ltd., P.R. China
%Shenzhen 518129, P.R. China, e-mail: zhuhufei@huawei.com and binli@huawei.com.
% and Huawei Technologies Co., Ltd., P.R. China
}% <-this % stops a space
%\thanks{W. Chen is with }
%\thanks{ is with the Department
%of Wireless Research, Huawei Technologies Co., Ltd.,
%Bantian, Longgang District, Shenzhen 518129, P.R.China, e-mail: }% <-this % stops a space
}

\markboth{XXXXXX}%
{Shell \MakeLowercase{\textit{et al.}}: Bare Demo of IEEEtran.cls for Journals}

\maketitle

\begin{abstract}
In this comment, we justify that the computational complexity
proposed in the paper "A New ML Based Interference Cancellation
Technique for Layered Space-Time Codes" (\emph{IEEE Trans. on
Communications}, vol.~57, no.~4, pp. 930-936, 2009) is $O(N^3)$
rather than the claimed $O(N^2)$, where $N$ is the number of receive
antennas.
%Partially inspired by the recently proposed recursive algorithm for vertical Bell Laboratories layered space-time
%architecture (V-BLAST), in this paper we propose a recursive algorithm for Groupwise Space-Time Block Coded system (G-STBC),
%which exploits %the properties of
%the Alamouti structure in the equivalent channel model to reduce the
%computational complexity dramatically. With respect to the existing
%efficient algorithms for G-STBC and those only for double space-time
%transmit diversity (DSTTD), the proposed minimum mean-square error
%(MMSE) ordered successive interference cancellation (OSIC) algorithm
%requires less complexity and achieves better performance. The
%proposed algorithm speeds up the MMSE sub-optimal OSIC algorithm for
%G-STBC by $2.57$, and speeds up the recently proposed DSTTD
%algorithm by up to $7.27$. Moreover, we propose improved
%implementations for the recursive V-BLAST algorithm and the proposed
%G-STBC algorithm, which save memories and permutation operations,
%and do not increase the complexity.
\end{abstract}

%\begin{IEEEkeywords}
%Fast recursive algorithms, MIMO, V-BLAST, G-STBC, DSTTD, Alamouti structure.
%\end{IEEEkeywords}

\IEEEpeerreviewmaketitle

%\section{Introduction}

%would like to
%$\bolddelta$
A maximum likelihood (ML) based interference cancellation (IC) detector
was proposed in \cite{1} for
%layered space time block codes (STBC)
double space-time transmit diversity (DSTTD),  %where DSTTD
which consists of two Alamouti's space-time block codes (STBC)
units \cite{fAlamoutiSTBC}. In many application areas of interest,
the computational complexity of the detector in \cite{1}
can be less than that of the conventional minimum
mean squared error (MMSE) IC detector for DSTTD \cite{SMSTBC2}. However,
%the number of computations (real multiplications and
%real additions)
the complexity claimed in \cite{1} needs to be modified,  as will be discussed
%shown
in this comment.
%We make some comments on the %computational
%complexity of

%your correction/addendum to the previous paper.
Let $N$ denote the number of receive antennas. In \cite{1}, the
theoretical analysis gaves a complexity of $O(N^2)$ (i.e.
$7N^2+62N-103$ real multiplications and $12N^2+47N-103$ real
additions) \cite[Table \Rmnum{1}]{1}, while numerical experiments
were not carried out to verify the given
complexity. %In this comment,
In what follows,
we show that the complexity is not $O(N^2)$, but
$O(N^3)$, and then %we
give the exact complexity that is verified by
our numerical experiments.
%To verify  by
%our numerical experiments to count the floating-point operations
%(flops).
%We show that the complexity is not $O(N^2)$, but
%$O(N^3)$. Then we give the exact complexity, which is verified by
%our numerical experiments to count the floating-point operations
%(flops).

Firstly, we show that a complexity of $O(N^3)$ is required
to perform the orthonormalization process by equations (9), (13),
(14) and (15) in \cite{1}. Let $(\bullet)^T$ and $ ( \bullet )^H $
denote transpose and conjugate transpose of a vector,
%matrix transposition and matrix conjugate transposition,
respectively. Equation (9) in \cite{1} defines the basis vectors
  \begin{equation}\label{eq9ofabvPp}
{\bf{v}}_i =\left[ {\begin{array}{*{20}c}
   {a_i} & {b_i} & {\bf e}_i^T  \\
\end{array}} \right]^T,
 \end{equation}
 where $i=1,2,\cdots,2N-2$, and ${\bf e}_i$ is the $(2N-2)\times 1$ vector with the $i^{th}$ element to be $1$
 and all others to be zero. Equation (13) in \cite{1} utilizes
%  ${\bf{v}}_i$ for , i.e.,
${\bf{v}}_1$ and ${\bf{v}}_2$, which is
%to obtain
\begin{equation}\label{eq13ofabvPp}
{\bm{\uptheta}}_{1}={\bf{v}}_1/\left\|  {\bf{v}}_1 \right\|,
 \quad {\bm{\uptheta}}_{2}={\bf{v}}_2/\left\|  {\bf{v}}_2 \right\|.
\end{equation}
%which is .
%is not
%the claimed $O(N^2)$ but $O(N^3)$.
Moreover, we represent equations (14) and (15) in \cite{1} as
\begin{subnumcases}{\label{EqTrnCm9NotPf1415}}
{\bm{\uptheta}}_{2n - 1}=\left( {{\bf{v}}_{2n - 1}  -
\sum\nolimits_{j = 1}^{2n - 2} {c_{2n-1}^j{\bm{\uptheta}}_j } }
\right)/\left\|  \cdots
\right\|, & \label{EquTranCom9NotPf14}\\
{\bm{\uptheta}}_{2n}  = \left( {{\bf{v}}_{2n}  - \sum\nolimits_{j =
1}^{2n - 2} {c_{2n}^j {\bm{\uptheta}}_j } } \right)/\left\|  \cdots
\right\|, &  \label{EquTranCom9NotPf15}
\end{subnumcases}
where
\begin{subnumcases}{\label{eq2}}
{c_{2n-1}^j={{\bm{\uptheta}}_j^H {\bf{v}}_{2n -
1}}}, & \label{eq2a}\\
{c_{2n}^j={{\bm{\uptheta}}_j^H {\bf{v}}_{2n}}}, &  \label{eq2b}
\end{subnumcases}
%In (\ref{EqTrnCm9NotPf1415}) and (\ref{eq2}),
and $n=2,3,\cdots,N-1$. It
can be seen that $\left[ {\begin{array}{*{20}c}
   {{\bm{\uptheta}}_{2n - 1} } & {{\bm{\uptheta}}_{2n} }  \\
\end{array}} \right]$ consists of $2\times2$ Alamouti sub-blocks \cite{Invariant_Alamouti}.
Thus we can obtain ${\bm{\uptheta}}_{2n}$ from
${\bm{\uptheta}}_{2n-1}$, to avoid computing
(\ref{EquTranCom9NotPf15}) and (\ref{eq2b}).
%, since ${\bm{\uptheta}}_{2n}$ can
%be obtained from ${\bm{\uptheta}}_{2n - 1}$ directly.

Let ${\bm{\uptheta}} \sim \left\lfloor {i,j, \cdots ,k}
\right\rfloor$
  denote that only the $i^{th},j^{th},\cdots,k^{th}$ entries in the vector ${\bm{\uptheta}}$ are non-zero.
  From (\ref{eq9ofabvPp}), we obtain
  \begin{equation}\label{eq3}
{\bf{v}}_i \sim \left\lfloor {1,2,i+2} \right\rfloor,
 \end{equation}
  where $i=1,2,\cdots,2N-2$. From (\ref{eq13ofabvPp}) and (\ref{eq3}), we
  obtain
  \begin{equation}\label{eq4}
{\bm{\uptheta}}_1 \sim \left\lfloor {1,2,3} \right\rfloor, \quad
{\bm{\uptheta}}_2 \sim \left\lfloor {1,2,4} \right\rfloor.
 \end{equation}
%and
%  \begin{equation}\label{eq5}
%{\bm{\uptheta}}_2 \sim \left\lfloor {1,2,4} \right\rfloor.
% \end{equation}
Let $n=2$ in (\ref{EqTrnCm9NotPf1415}) to obtain
  \begin{multline}\label{eq6}
{\bm{\uptheta}}_3 = \left( {{\bf{v}}_3  - c_3^1 {\bm{\uptheta}}_1  -
c_3^2 {\bm{\uptheta}}_2 } \right)/\left\|  \cdots  \right\| \\ \sim
\left\lfloor {1,2,3,4,5} \right\rfloor=\left\lfloor {1-5}
\right\rfloor
 \end{multline}
 and
  \begin{equation}\label{eq7}
{\bm{\uptheta}}_4 = \left( {{\bf{v}}_4  - c_4^1 {\bm{\uptheta}}_1 -
c_4^2 {\bm{\uptheta}}_2 } \right)/\left\| \cdots \right\| \sim
\left\lfloor {1-4,6} \right\rfloor,
 \end{equation}
 where (\ref{eq3}) and (\ref{eq4})
 are  utilized.
% Then let $n=3$ in (\ref{EqTrnCm9NotPf1415}) to obtain
%  \begin{multline}\label{eq8}
%{\bm{\uptheta}}_5  = \left( {{\bf{v}}_5 - c_5^1 {\bm{\uptheta}}_1  -
%c_5^2 {\bm{\uptheta}}_2 - c_5^3 {\bm{\uptheta}}_3  - c_5^4
%{\bm{\uptheta}}_4 } \right)/\left\| \cdots \right\| \\ \sim
%\left\lfloor {1-7} \right\rfloor
% \end{multline}
% and
%  \begin{multline}\label{eq9}
%{\bm{\uptheta}}_6 = \left( {{\bf{v}}_6  - c_6^1 {\bm{\uptheta}}_1 -
%c_6^2 {\bm{\uptheta}}_2 - c_6^3 {\bm{\uptheta}}_3  - c_6^4
%{\bm{\uptheta}}_4 } \right)/\left\| \cdots \right\|\\ \sim
%\left\lfloor {1-6,8} \right\rfloor,
% \end{multline}
% where (\ref{eq3})$-$(\ref{eq7}) are utilized.
 From (\ref{eq4})$-$(\ref{eq7}),
 %(\ref{eq4,eq5,eq6,eq7,eq8,eq9}),
 it can be seen that for $n=1,2$, we have
  \begin{equation}\label{eq10}
{\bm{\uptheta}}_{2n-1} \sim \left\lfloor {1-(2n+1)} \right\rfloor,
\quad {\bm{\uptheta}}_{2n} \sim \left\lfloor {1-2n,2n+2}
\right\rfloor.
 \end{equation}

Assume for any $n$, ${\bm{\uptheta}}_{2n-1}$ and
${\bm{\uptheta}}_{2n}$ satisfy (\ref{eq10}). This assumption will be
verified in this paragraph. From (\ref{EqTrnCm9NotPf1415}), it can
be seen that ${\bm{\uptheta}}_{2(n+1)-1}$ includes the sum of
${\bm{\uptheta}}_{2n-1}$, ${\bm{\uptheta}}_{2n}$ and
${\bf{v}}_{2(n+1) - 1}$, while ${\bm{\uptheta}}_{2(n+1)}$ includes
the sum of ${\bm{\uptheta}}_{2n-1}$, ${\bm{\uptheta}}_{2n}$ and
${\bf{v}}_{2(n+1)}$. From (\ref{eq3}) and the assumption
(\ref{eq10}),
%${\bf{v}}_{2(n+1) - 1}$ and ${\bf{v}}_{2(n+1)}$
%satisfy (\ref{eq3}), while  ${\bm{\uptheta}}_{2n-1}$ and
%${\bm{\uptheta}}_{2n}$ satisfy (\ref{eq10}).
we can conclude that ${\bm{\uptheta}}_{2(n+1)-1}$ and
${\bm{\uptheta}}_{2(n+1)}$ also satisfy (\ref{eq10}). Then the
assumption (\ref{eq10}), which is valid for $n=1,2$, is still
%we can conclude that the assumption (\ref{eq10}) is
valid for all the subsequent $(n+1)$s where $n=1,2,\cdots,N-2$. Thus
we have verified the assumption (\ref{eq10}) for any $n$.

It can be seen from (\ref{eq10}) that in (\ref{EquTranCom9NotPf14}),
$c_{2n-1}^j{\bm{\uptheta}}_j$ requires more than $j$
multiplications, while $\sum\nolimits_{j = 1}^{2n - 2}
{c_{2n-1}^j{\bm{\uptheta}}_j}$ requires more than $\sum\nolimits_{j
= 1}^{2n - 2} {j}\approx2n^2$ multiplications. Then totally it
requires more than $\sum\nolimits_{n = 2}^{N-1} {2n^2}\approx
\frac{2}{3}N^3$ multiplications to compute
(\ref{EquTranCom9NotPf14}) for $n=2,3,\cdots,N-1$. Thus we have
shown that the actual complexities of the detector in \cite{1}
should be at least $O(N^3)$.

\begin{table*}[!t]
\renewcommand{\arraystretch}{1.3}
\caption{The computational complexities of the equations in
\cite{1}} \label{table_III} \centering
\begin{tabular}{c|c|c|c|c|}
%& \multicolumn{3}{c|}{The Detector} & \multicolumn{3}{c|}{The Proposed Detector for DSTTD} \\
%& \multicolumn{3}{c|}{for DSTTD in \cite{Jun09add_TransCom}} & \multicolumn{3}{c|}{(max. mean value$\sim$min. mean value)} \\
%\hline
%\bfseries      & \bfseries  Real & \bfseries Real & \bfseries  Total    & \bfseries Real & \bfseries Real & \bfseries Total Flops \\
%\bfseries   N  & \bfseries  Mult. & \bfseries Add. & \bfseries Flops  & \bfseries Mult. & \bfseries Add. & \bfseries $\frac{16}{3}N^3+69N^2+\frac{37}{3}N-25$  \\
%\hline
\bfseries  Equation Number  &\bfseries Complex Multiplications  &\bfseries Complex Additions &\bfseries  Real Multiplications  &\bfseries Real Additions  \\
\hline
\bfseries  (9) and (11)  &4(N-1) &2(N-1) &4(N-1)+4 &3   \\
\hline
\bfseries  (13)  & &  &9   &4      \\
\hline
\bfseries  (14)  &$\frac{2}{3}N(N-1)(N-2)$ &$\frac{2}{3}N(N-1)(N-2)$  &$(6N+5)(N-2)$   &$2(N+1)(N-2)$     \\
\hline
\bfseries  (23)  &$2(N-1)N$  &$2(N-1)N$  &$4(N-1)$   &     \\
\hline
\bfseries  (25)  &$2(N-1)N$  &$2(N-1)N$  &$4(N-1)$   &    \\
\hline
\bfseries  (28)  &$4N$  &$4N$  &   &    \\
\hline
\bfseries  Sum  &$\frac{2}{3}N^3+2N^2+\frac{16}{3}N-4$  &$\frac{2}{3}N^3+2N^2+\frac{10}{3}N-2$  &$6N^2+5N-9$   &$2N^2-2N+3$    \\
\hline
\end{tabular}
\end{table*}

%which are

The dominant computations of the ML based IC detector \cite{1} come
from equations (9), (11), (13), (14), (23), (25) and (28) in
\cite{1}, of which the complexities are listed in Table \Rmnum{1}.
%we list  of the equations in \cite{1}.
One complex multiplication takes four real multiplications and two
real additions, while one complex addition needs two real additions.
Therefore, it can be seen from Table \Rmnum{1} that the complexities
of the detector are equivalent to
\begin{equation}\label{eqRealmulti}
\frac{8}{3}N^3+14N^2+\frac{79}{3}N-25
\end{equation}
real multiplications and
\begin{equation}\label{eqRealAdd}
\frac{8}{3}N^3+10N^2+\frac{46}{3}N-9
\end{equation}
real additions. The total complexity is the sum of real
multiplications and additions \cite{1}, which is
%In a digital signal processor (DSP) implementation, both addition
%and multiplication can be done in one clock cycle.  counts in DSP
%implementation. The total computational complexity of the proposed
%algorithm (Cp) is Cp = 19N2 + 109N . 206, whereas that of the MMSE
%IC algorithm (Cm) is Cm = 15N3 + 73 2 N2 . 3 2N. Note that Cp < Cm
%regardless of N regime, which clearly shows the advantage of the
%proposed algorithm in terms of complexity. The complexity of the
%proposed algorithm can be about 5 times smaller than that of the
%MMSE IC algorithm when the number of Rx antennas is 8. It can be
%observed that the Correspondingly the detector in \cite{1} requires
%\begin{multline}\label{eq12}
%(\frac{8}{3}N^3+37N^2+\frac{37}{3}N-4)+(\frac{8}{3}N^3+32N^2+\frac{28}{3}N-21)
%\\ = \frac{16}{3}N^3+69N^2+\frac{37}{3}N-25
%\end{multline}
\begin{equation}\label{eq12}
\frac{16}{3}N^3+24N^2+\frac{125}{3}N-34
\end{equation}
floating-point operations
(flops). We also carried out numerical experiments to count the
flops
required by the detector in \cite{1}.
%, for $N=2,3,\cdots,8$.
% have yielded
The results of our numerical experiments  %that are
are identical to those computed by (\ref{eq12}), i.e., our numerical
experiments have accurately verified (\ref{eq12}).

\begin{table}[!t]
\renewcommand{\arraystretch}{1.3}
\caption{Complexity comparison} \label{table_III} \centering
\begin{tabular}{c|c|c|c|c|c|c|}
& \multicolumn{3}{c|}{The ML based IC} & \multicolumn{3}{c|}{The MMSE IC} \\
& \multicolumn{3}{c|}{detector for DSTTD \cite{1}} & \multicolumn{3}{c|}{detector for DSTTD \cite{SMSTBC2}} \\
\hline
      & \bfseries  Real & \bfseries Real & \bfseries  Total    & \bfseries Real & \bfseries Real & \bfseries Total \\
\bfseries   N  & \bfseries  Mult. & \bfseries Add. & \bfseries Flops  & \bfseries Mult. & \bfseries Add. & \bfseries Flops \\
\hline
\bfseries 2   &105 &83 &188  &128  &135   &263  \\
\hline
\bfseries 3  &252 &199  &451 &360   &369  &729  \\
\hline
\bfseries 4  &475 &383  &858   &768    &770  &1538  \\
\hline
\bfseries 5  &790 &651  &1441   &1400   &1380   &2780  \\
\hline
\bfseries 6  &1213  &1019  &2232   &2304    &2241  &4545  \\
\hline
\bfseries 7  &1760  &1503  &3263   &3528    &3395  &6923  \\
\hline
\bfseries 8  &2447  &2119  &4566   &5120    &4884  &10004  \\
\hline
\end{tabular}
\end{table}
Table \Rmnum{1} in \cite{1} compared the complexities of the ML
based IC detector for DSTTD in \cite{1} and the conventional MMSE IC
detector for DSTTD in \cite{SMSTBC2}.
From (\ref{eqRealmulti}), (\ref{eqRealAdd}) and (\ref{eq12}), it can
be seen that Table \Rmnum{1} in \cite{1} should be modified to Table
\Rmnum{2} in this comment, where the total complexity of the MMSE IC
detector in \cite{SMSTBC2} is
\begin{equation}\label{eqOldMMSEic}
15N^3+\frac{73}{2}N^2-\frac{3}{2}N
\end{equation}
flops \cite{1}. From Table \Rmnum{2}, it can be seen that the
complexity of the detector proposed in \cite{1} is about $2.2$ times
smaller than that of the MMSE IC detector \cite{SMSTBC2} when the
number of receive antennas is $8$.

\ifCLASSOPTIONcaptionsoff
  \newpage
\fi

%\begin{figure}[!t]
%\centering
%\includegraphics[width=6.5in]{trans1BLAST.eps}
%\caption{Comparison of computational complexities among the
%square-root MMSE-OSIC V-BLAST algorithms in \cite{zhf2},
%\cite{vtc08_zhfVTC4} and this paper, the recursive MMSE-OSIC V-BLAST
%algorithms in \cite{zhf3,reviewer_VTC08_b,zhf6 }, and the ZF-OSIC
%V-BLAST algorithm
% in \cite{vtc08_reviewer_a}. In the legend of this figure, ``sqrt" and ``Alg" are the
%abbreviations for square-root and algorithm, respectively, while
%``sqrt Alg $\cdots$ with Householder" and ``sqrt Alg $\cdots$ with
%Givens" block upper-triangularize the square-root matrix by a
%householder reflection and a sequence of Givens rotations,
%respectively.} \label{fig_sim}
%\end{figure}

%\newpage

\end{document}